\documentstyle[11pt,aaspp4,flushrt]{article}
\def \beq{\begin{equation}}
\def \eeq{\end{equation}}
\def \IGM{{\rm IGM}}
\def\lya{Ly$\alpha$~}
\def\zr{z_{\rm reion}}
\begin{document}

\title{Identifying the Reionization Redshift from the Cosmic Star Formation
Rate}

\author{Rennan Barkana\altaffilmark{1} and Abraham Loeb\altaffilmark{2}}

\altaffiltext{1}{Institute for Advanced Study, Olden Lane, Princeton,
NJ 08540; email: barkana@ias.edu}

\altaffiltext{2}{Harvard-Smithsonian Center for Astrophysics, 60 Garden St.,
Cambridge, MA 02138; email: aloeb@cfa.harvard.edu}

\begin{abstract}

We show that the cosmic star formation rate per comoving volume should
exhibit a distinct drop around the reionization redshift, when the
\ion{H}{2} regions in the intergalactic medium around individual
ionizing sources first overlapped. The drop results from the increase
in the temperature of the intergalactic medium as it was
photo-ionized, and the consequent suppression of the formation of
low-mass galaxies. We show quantitatively that the detection of this
drop, which is marked by a corresponding fall in the number counts of
faint galaxies, should become feasible over the coming decade with the
{\it Next Generation Space Telescope}.

\end{abstract}


\centerline{Submitted to {\it The Astrophysical Journal}, 2000}

\section{Introduction}

Current observations reveal the existence of galaxies out to redshifts
as high as $z\sim 6.7$ (Chen et al.\ 1999; Weymann et al.\ 1998; Dey
et al.\ 1998; Spinrad et al.\ 1998; Hu, Cowie, \& McMahon 1998) and
bright quasars out to $z\sim 5$ (Fan et al.\ 1999). The detection of
transmitted flux at rest-frame wavelengths shorter than \lya in the
spectrum of some of these sources implies that most of the
intergalactic medium (IGM) is ionized at $z\sim 5$ (Songaila et al.\
1999), and that a substantial population of ionizing sources should
exist at higher redshifts (Madau, Haardt, \& Rees 1999). This
inference is consistent with theoretical expectations in Cold Dark
Matter (CDM) models of galaxy formation (Shapiro, Giroux, \& Babul
1994; Gnedin \& Ostriker 1997; Haiman \& Loeb 1998a,b; Valageas \&
Silk 1999; Miralda-Escud\'e, Haehnelt, \& Rees 1999; Chiu \& Ostriker
1999; Gnedin 1999).

One of the major goals of the study of galaxy formation is to achieve
an observational determination and a theoretical understanding of the
cosmic star formation history. By now, this history has been sketched
out only to a redshift $z\sim 4$ (e.g., see the compilation of Blain
et al.\ 1999). The {\it Next Generation Space Telescope 
(NGST}\,)\footnote{See http://www.ngst.stsci.edu/},
planned for launch in 2008, is expected to reach an imaging
sensitivity better than 1 nJy in the infrared, which will allow it to
detect galaxies and hence determine the star formation history at
$z\ga 10$.

The redshift of reionization, $\zr$, defines the epoch when the \ion{H}{2}
regions surrounding individual sources in the IGM overlapped. At this time,
the cosmic ionizing background increased sharply, and the IGM was heated by
the ionizing radiation to a temperature $\ga 10^4$ K (see, e.g., Gnedin
1999). In this {\it Letter}, we explore the distinct signature that this
heating process is expected to have left on the global star formation rate
in the universe. Due to the substantial increase in the IGM temperature,
the intergalactic Jeans mass increased dramatically, changing the minimum
mass of forming galaxies (Gnedin \& Ostriker 1997; Miralda-Escud\'e \& Rees
1998). Reionization is therefore predicted to cause a drop in the cosmic
star formation rate (SFR) at $\zr$.  This drop is accompanied by a dramatic
fall in the number counts of faint galaxies. Detection in future galaxy
surveys of this fall in the faint luminosity function could be used to
identify $\zr$ observationally. This method for inferring $\zr$ is more
straightforward than alternative proposals which rely on high resolution
spectroscopy of individual sources (Miralda-Escud\'e 1998; Haiman \& Loeb
1999; Loeb \& Rybicki 1999) or on the detection of faint spectral features
in the background radiation (Gnedin \& Ostriker 1997; Shaver et al.\ 1999).

In \S 2 we describe the galaxy formation model used to calculate the cosmic
SFR and the luminosity function at high redshift. The results from this
model, including the reionization signature, are described in \S 3. We
dedicate \S 4 to a detailed comparison with previous semi-analytic models
and numerical simulations.  Finally, we summarize our main conclusions in
\S 5.

\section{Galaxy Formation Model}

We model galaxy formation within a hierarchical CDM model of structure
formation. We obtain the abundance of dark matter halos using the
Press-Schechter (1974) model. Relevant expressions for
various CDM cosmologies are given, e.g., in Navarro, Frenk, \& White
(1997). The abundance of halos evolves with redshift as each halo
gains mass through mergers with other halos. If $dp[M_1,t_1
\rightarrow M,t]$ is the probability that a halo of mass $M_1$ at time
$t_1$ will have merged to form a halo of mass between $M$ and $M+dM$
at time $t>t_1$, then in the limit where $t_1$ tends to $t$ we obtain
an instantaneous merger rate $d^2p[M_1\rightarrow M,t]/(dM\ dt)$. This
quantity was evaluated by Lacey \& Cole [1993, their Eq.\ (2.18)].

Once a dark matter halo has collapsed and virialized, the two
requirements for forming new stars are gas infall and cooling.  Gas of
primordial composition can cool through atomic transitions in halos
with virial temperatures $T_{\rm vir}\ga 10^{3.8}$K (Haiman, Abel, \&
Rees 1999), and through molecular hydrogen (${\rm H_2}$) transitions
in smaller halos. However, ${\rm H_2}$ molecules are fragile, and
could have been easily photo-dissociated throughout the Universe by
trace amounts of starlight (Stecher \& Williams 1967; Haiman, Rees, \&
Loeb 1996) that were well below the level required for complete
reionization of the IGM. We therefore assume that, in the redshift
interval covered by our models, efficient cooling can only occur with
atomic transitions, in halos above a circular velocity of $V_c\sim
13.3\ {\rm km\ s}^{-1}$. Haiman et al.\ (1999) showed that if quasars
contributed significantly to the UV background before reionization,
their soft X-rays produced free electrons in small halos and may have
catalyzed the formation of molecular hydrogen. In this case, smaller
halos than we assume could have formed stars prior to reionization,
down to $T_{\rm vir}\ga 10^{2.4}$K. However, the same X-rays may also
have begun to heat the IGM before reionization, hindering gas infall
into the smallest halos. Even if this heating mechanism was
ineffective, most of the additional galaxies would be too faint to
detect with {\it NGST.}\, Regardless of the effectiveness of $H_2$
cooling, the formation of new galaxies after reionization would still
be greatly suppressed since it is limited by gas infall.

Gas infall depends sensitively on the Jeans mass. When a halo more
massive than the Jeans mass begins to form, the gravity of its dark
matter overcomes the gas pressure. Even in halos below the Jeans mass,
although the gas is initially held up by pressure, once the dark
matter collapses its increased gravity pulls in some gas (Haiman,
Thoul, \& Loeb 1997). Before reionization, the IGM is cold and
neutral, and the Jeans mass plays a secondary role compared to
cooling. After reionization, the Jeans mass is increased by several
orders of magnitude due to the photoionization heating of the IGM, and
hence begins to play a dominant role in limiting the formation of
stars. Gas infall in a reionized and heated Universe has been
investigated in a number of numerical simulations. Thoul \& Weinberg
(1996) found a reduction of $\sim 50\%$ in the collapsed gas mass due
to heating, for a halo of circular velocity $V_c\sim 50\ {\rm km\
s}^{-1}$ at $z=2$, and a complete suppression of infall below $V_c
\sim 30\ {\rm km\ s}^{-1}$. Kitayama \& Ikeuchi (2000) also performed
spherically-symmetric simulations but they included self-shielding of
the gas, and found that it lowers the circular velocity thresholds by
$\sim 5\ {\rm km\ s}^{-1}$.  Three dimensional numerical simulations
(Quinn, Katz, \& Efstathiou 1996; Weinberg, Hernquist, \& Katz 1997;
Navarro \& Steinmetz 1997) found a significant suppression of gas
infall in even larger halos ($V_c \sim 75\ {\rm km\ s}^{-1}$), but
this was mostly due to a suppression of late infall at $z\la 2$. Thus,
we adopt a prescription for the suppression of gas infall based on the
spherically symmetric simulations. Each of these simulations follows
an isolated halo, rather than a set of merging progenitors, and the
halo is followed for longer than it would typically survive before
merging into a larger halo. Therefore, we adopt slightly higher $V_c$
thresholds than indicated by the spherically symmetric simulations.

When a volume of the IGM is ionized by stars, the gas is heated to a
temperature $T_{\IGM}\sim 10^4$ K. If quasars dominate the UV
background at reionization, their harder photon spectrum leads to
$T_{\IGM}\sim 2\times 10^4$ K. Including the effects of dark matter, a
given temperature results in a linear Jeans mass (e.g., \S 6 in
Peebles 1993) corresponding to a halo circular velocity of \beq V_J=93
\left(\frac{T_{\IGM}}{2\times 10^4 {\rm K}}\right)^{1/2}\
\left[\frac{1}{\Omega(z)}\ \frac{\Delta_c}{18 \pi^2}\right]^{1/6}\
{\rm km\ s}^{-1}, \eeq where $\Delta_c$ is the mean collapse
overdensity (Bryan \& Norman 1998) and $\Omega(z)$ is the total
density in units of the critical density, evaluated at redshift
$z$. In halos with $V_c>V_J$, the gas fraction equals the universal
mean of $\Omega_b/\Omega_0$. We assume that the gas fraction is
suppressed by $50\%$ at a circular velocity $V_c=V_J\times 55/93$ and
a complete suppression occurs below $V_c=V_J\times 35/93$. We set
$T_{\IGM}= 2\times 10^4$ K, although a lower temperature would result
in a slightly weaker suppression of gas infall.

Recent semi-analytic models (Miralda-Escud\'e et al.\ 1999) and
numerical simulations (Gnedin 1999) show that the process of
reionization is gradual. When stars or quasars form in galaxies, each
source must first ionize gas in its host object (Wood \& Loeb
2000). Individual ionization bubbles then grow for about a Hubble time
at the relevant redshifts, until they begin to overlap.  Since
overlapping \ion{H}{2} regions are filled with radiation from several
sources, the ionizing intensity increases rapidly at each overlap and
this leads to a quick ionization of almost the entire volume of the
universe. Only the highest density gas remains neutral and is
gradually ionized subsequently.

The reionization feedback on 
galaxy formation depends on the fraction of the
IGM which is ionized at each redshift. Thus, we focus on the overlap
stage of reionization. We refer to the reionization redshift $\zr$
as the redshift at which $50\%$ of the volume of the universe is
ionized, and assume that a significant volume is first ionized at the
starting redshift $z_{\rm start}$ and that most of the volume has been
ionized by the ending redshift $z_{\rm end}$. We take the Gnedin
(1999) simulation as a guide for the redshift interval of
reionization, adopting $z_{\rm start}= 11$ and $z_{\rm end}= 6.5$ for
the case of $\zr = 8.2$. When we vary $\zr$ we scale the values of
$z_{\rm start}$ and $z_{\rm end}$ in order to keep the reionization
era equal to a constant fraction of the age of the universe at $\zr$.

At each redshift, we thus obtain the fraction of the Universe which is
ionized, and, in both the neutral and ionized IGM, we obtain the gas
mass which falls and cools into halos as a function of the halo mass
$M$. If the cold gas mass is denoted by $f_{\rm cold}(M,z)\, M
\Omega_b/\Omega_0$, then $f_{\rm cold}(M,z)=0$ in halos below a minimum
mass $M_{\rm min}(z)$, set by the minimum circular velocity. Given
$f_{\rm cold}(M,z)$, we then derive the SFR  
directly from halo mergers,
and the luminosity of each galaxy directly from its SFR.

New star formation in a given galaxy can occur either from primordial
gas or from recycled gas which has already undergone a previous burst
of star formation. Consider first a region with neutral IGM, in which
$M_{\rm min}(z)$ is set by the need for gas to cool. In this case,
halos below $M_{\rm min}(z)$ contain roughly the universal fraction
$\Omega_b/\Omega_0$ of primordial gas, which collapsed along with the
dark matter but could not cool. Halos with $M>M_{\rm min}(z)$ contain
the same gas fraction, but this gas {\em is} able to cool [i.e.,
$f_{\rm cold}(M,z)=1$]. Consider halos of mass $M_1$ and $M_2$ merging
to yield a halo of total mass $M$. We assume that the merger triggers
star formation only if $M>M_{\rm min}(z)$. If $M_1 < M_{\rm min}(z)$,
then the merger will trigger star formation in the primordial gas
contained in this halo. Assuming a star formation efficiency $\eta$,
we obtain a starburst with a total stellar mass of $\eta M_1
\Omega_b/\Omega_0$. Similarly, if $M_2 < M_{\rm min}(z)$ then the
starburst has an additional contribution equal to a mass of $\eta M_2
\Omega_b/\Omega_0$. If $M_1 > M_{\rm min}(z)$, we assume that the gas
in this halo has already undergone a previous burst of star formation,
and thus, before the merger, the gas fraction is only
$(1-\eta)\times\Omega_b/\Omega_0$.  We assume that the merger triggers
another episode of star formation in this gas if $M_2$ is sufficiently
massive. Numerical simulations of starbursts in interacting $z=0$
galaxies (e.g., Mihos \& Hernquist 1994; 1996) found that a merger
triggers significant star formation even if $M_2$ is a small fraction
of $M_1$.  Preliminary results (Somerville, private communication)
from simulations of mergers at $z \sim 3$ find that they remain
effective at triggering star formation even when the initial disks are
dominated by gas. We adopt a conservative threshold mass ratio of
$r_{\rm merge}=1/2$ and assume that if $M_2> r_{\rm merge} M_1$ then
star formation is triggered in $M_1$, producing a stellar mass of
$\eta (1-\eta) M_1 \Omega_b/\Omega_0$ (where we take the same
efficiency, $\eta$, for star formation from primordial and from
recycled gas). Similarly, if $M_2 > M_{\rm min}(z)$, then a stellar
mass of $\eta (1-\eta) M_2 \Omega_b/\Omega_0$ is produced as long as
$M_1 > r_{\rm merge} M_2$.

Star formation proceeds along similar lines in a region with ionized IGM,
except that the value of $M_{\rm min}(z)$, the lowest mass of a halo which
can accumulate gas from the IGM, is set by gas infall. An additional
complication in this case is that some galaxies in halos below $M_{\rm
min}(z)$ could have survived photo-ionization heating (Barkana \& Loeb
1999a) and kept their cold gas reservoir after reionization. Although gas
infall into these galaxies was subsequently suppressed, they may have
continued to merge and thus form stars. In order to be conservative about
the suppression of star formation, we assume that all gas which cooled
before reionization settled into dense gas disks and did not
photo-evaporate subsequently. We estimate the contribution of this gas to
merger-induced star formation as follows.  At a given redshift $z$ after
reionization has started, we find the redshift $z_{\rm half}$ at which the
filling factor of ionized regions was equal to half its value at $z$. Thus,
roughly half of the volume which is ionized at $z$ was ionized after
$z_{\rm half}$. We compute the gas fraction $f_{\rm coll}$ which had
collapsed into halos in neutral regions by redshift $z_{\rm half}$. This
gas lies in halos that will have merged into larger halos by redshift
$z$. To approximate the resulting gas content of halos, we assume that at
redshift $z$ all halos down to a mass $M_{\rm merge}(z)$ contain a gas
fraction equal to the cosmic mean of $\Omega_b/\Omega_0$, where $M_{\rm
merge}(z)$ is set by requiring the total gas fraction in these halos to
equal $f_{\rm coll}$. Since $M_{\rm merge}(z)$ may be lower than $M_{\rm
min}(z)$, we include by this prescription halos which had formed prior to
reionization but could not have begun to form in a fully ionized universe.
As discussed above, if we only considered infall in an ionized region then
halos with $M<M_{\rm min}(z)$ would contain no gas, and the gas fraction
would increase gradually up to $f_{\rm cold}(M,z)=1$ for all $M$ greater
than the Jeans mass at $z$. We modify this prescription by increasing
$f_{\rm cold}(M,z)$ to unity whenever $M>M_{\rm merge}(z)$. Thus, when two
halos with masses $M_1$ and $M_2=M-M_1$ merge, the resulting halo accretes
from the IGM an additional primordial gas mass of $[f_{\rm
cold}(M,z)M-f_{\rm cold}(M_1,z)M_1-f_{\rm cold}(M_2,z)M_2]
[\Omega_b/\Omega_0]$ and turns a fraction $\eta$ of it into stars. In
addition, the initial stellar content of the two halos may undergo
merger-induced recycled star formation, as discussed above in the
pre-reionization case.

In order to determine the typical SFR in a halo of mass $M$, we average the
SFR over all halos $M_1<M$ that can merge to produce a halo of fixed total
mass $M$. We use the Press-Schechter (1974) abundance of halos of mass
$M_1$, and the halo merger rate $d^2p[M_1\rightarrow M,t]/(dM\ dt)$
introduced above. By averaging the SFR over all possible mergers we neglect
the possibility of episodic star formation in individual galaxies, with
merger-triggered peaks reaching star formation rates well above the
average. However, for the halo mass scales relevant to galaxy formation at
high redshift, the slope of the primordial power spectrum approaches
$n=-3$, which implies that the merger rate is high and galaxies double
their mass in a short time compared to the age of the universe. Galaxies
can be brighter than we calculate only if they undergo star formation on a
timescale much shorter than the merger timescale. We expect that at each
redshift, galaxies observed during an exceptional burst of star formation
would, on the whole, account for only a small fraction of the total cosmic
SFR.

Having obtained the average SFR for a halo of mass $M$, we calculate
the luminosity of this halo self-consistently from this SFR. At a
given redshift, we assume that the halo contains a total stellar mass
of $\eta f_{\rm cold}(M,z) M \Omega_b/\Omega_0$ which has formed at a
steady rate equal to the calculated average SFR. The total stellar
mass in each halo may be slightly different from a fraction $\eta$ of
the cold gas mass, but the luminosity depends mostly on the number of
young, bright stars, and this number depends only on the SFR at the
time that the galaxy is observed.

\section{Results}

We assume a $\Lambda$CDM cosmology, with matter and vacuum density
parameters $\Omega_0=0.3$ and $\Omega_{\Lambda}=0.7$, respectively. We
also assume a Hubble constant $H_0=70\mbox{ km
s}^{-1}\mbox{Mpc}^{-1}$, and a primordial scale invariant ($n=1$)
power spectrum with $\sigma_8=0.9$, where $\sigma_8$ is the
root-mean-square amplitude of mass fluctuations in spheres of radius
$8\ h^{-1}$ Mpc. We adopt a cosmological baryon density of
$\Omega_b=0.04$. 

In calculating the stellar emission spectrum, we assume a universal
Salpeter initial mass function with a metallicity $Z=0.001$, and use
the stellar population model results of Leitherer et al.\ (1999)
\footnote{Model spectra of star-forming galaxies were obtained from
http://www.stsci.edu/science/starburst99/}. We also include a
Ly$\alpha$ cutoff in the spectrum due to absorption by the dense
Ly$\alpha$ forest.  We do not, however, include dust extinction, which
is expected to be less significant at high redshift, when the mean
metallicity was low. We describe the sensitivity of {\it NGST} by
$F_{\nu}^{\rm ps}$, the minimum spectral flux at wavelengths
0.6--3.5$\mu$m required to detect a point source. The detection
threshold of extended sources is higher, and we therefore incorporate
a probability distribution of disk sizes at each value of halo mass
and redshift (see Barkana \& Loeb 1999b for details). We adopt a value
of $F_{\nu}^{\rm ps}=0.25$ nJy
\footnote{We obtained the flux limit using the {\it NGST} calculator
at http://www.ngst.stsci.edu/nms/main/}, assuming a very deep 300-hour
integration on an 8-meter {\it NGST} and a spectral resolution of
10:1.  This resolution should suffice for a $\sim 10\%$ redshift
measurement, based on the Ly$\alpha$ cutoff.

Figure 1 shows our predictions for the star formation history of the
Universe, with $\eta=10\%$. We show the SFR for $\zr=7$ (solid curves),
$\zr=10$ (dashed curves), and $\zr=13$ (dotted curves). In each pair of
curves, the upper one is the total SFR, and the lower one is the fraction
detectable with {\it NGST}. As noted in \S 2, if star formation is episodic
then halos are much brighter than average a small fraction of the time, and
the detectable SFR could be somewhat higher than shown in the figure,
although it would always lie below the total SFR.  As discussed above,
photoionization directly suppresses new gas infall after reionization, but
it does not immediately affect mergers which continue to trigger star
formation in gas which had cooled prior to reionization. Indeed, we find a
stronger suppression of primordial star formation (which declines by a
factor of 3.4 for $\zr=10$) than recycled star formation (which only
declines by a factor of 1.7 for $\zr=10$). The contribution from
merger-induced star formation is comparable to that from primordial gas at
$z<\zr$, and it is smaller at $z>\zr$. However, the recycled gas
contribution to the {\em detectable} SFR is dominant at the highest
redshifts, since infalling halos more massive than $M_{\rm min}(z)$
dominate the star formation in the most massive halos and only these most
massive halos are bright enough to be detected from the pre-reionization
era. Although most stars at $z\ga \zr$ form out of primordial,
zero-metallicity gas, a majority of stars in detectable galaxies may form
out of the small gas fraction that has already been enriched by the first
generation of stars.

Points with error bars in Figure 1 are observational estimates of the
cosmic SFR per comoving volume at various redshifts (see Blain et al.\
1999 for the original references). The highest SFR estimates at $z\sim
3$--4 are based on sub-millimeter observations or on
extinction-corrected observations at shorter wavelengths, and all are
fairly uncertain. We choose $\eta=10\%$ to obtain a rough agreement
between the models and these observations. The SFR curves are roughly
proportional to the value of $\eta$. Note that in reality $\eta$
may depend on the halo mass, since the effect of supernova feedback
may be more pronounced in small galaxies. 

Figure 1 shows a sharp rise in the total SFR at redshifts higher than
$\zr$.  Although only a fraction of the total SFR can be detected with {\it
NGST}, the detectable SFR displays a definite signature of the reionization
redshift. As noted in \S 2, if quasars were abundant before reionization
then stars may have formed in smaller halos through ${\rm H_2}$
cooling\footnote{Note, however, that the harder quasar spectra may lead to
broader ionization fronts and to pre-heating of the neutral IGM that would
suppress gas infall into the lowest-mass halos. This negative effect was not
considered by Haiman et al.\ (1999).}  (Haiman et al.\ 1999).  In this
case, the SFR rise would be even larger than in the cases shown in Figure
1, but the detected SFR would be essentially unchanged since the additional
galaxies would be extremely faint.

Most of the increase in SFR beyond the reionization redshift is due to star
formation occurring in very small, and thus faint, galaxies.  This
evolution in the faint luminosity function constitutes the clearest
observational signature of the suppression of star formation after
reionization. Figure 2 shows the predicted luminosity function of galaxies
at various redshifts. The curves show $d^2N/(dz\ d\log F_{\nu}^{\rm ps})$,
where $N$ is the total number of galaxies in a single field of view of {\it
NGST}. Results are shown at redshift $z=7$ (solid curves) or $z=13$ (dashed
curves).  In each pair of curves, the upper one at 0.1 nJy assumes a
permanently neutral IGM (i.e., $\zr \ll z$), and the lower one assumes a
permanently fully-ionized IGM (i.e., $\zr \gg z$). Although our models
assign a fixed, average, luminosity to all halos of a given mass and
redshift, in reality such halos would have some dispersion in their merger
histories and thus in their luminosities. We thus include smoothing in the
plotted luminosity functions, assuming that the flux of each halo can vary
by up to a factor of 2 around the mean for the set of halos with its mass
and redshift. As noted in \S 2, if star formation is episodic than the
distribution of fluxes about the mean could have a significant tail toward
high luminosities. Note the enormous increase in the number density of
faint galaxies in a pre-reionization universe. Observing this dramatic
increase toward high redshift would constitute a clear detection of
reionization and of its major effect on galaxy formation. The effect is
greatest below a flux limit of 1 nJy, and detecting it therefore requires
the capabilities of an 8-meter {\it NGST}.\, In the case of effective ${\rm
H_2}$ cooling before reionization, the luminosity functions in a neutral
IGM would follow the corresponding curves in Figure 2 down to their peaks,
but they would continue to rise toward even fainter fluxes. This additional
rise, though, would occur well below the detection limit of {\it NGST}\,.

\section{Comparison with previous work}

The reionization of the IGM has been explored in the past, based on
semi-analytic models as well as numerical simulations (see \S 1). In
this section we discuss the relation between our work and previous
models that considered the evolution of star formation through the
epoch of reionization.

The key physical ingredient which leads to the SFR drop after
reionization is the temperature gap between the warm ionized regions
(where the Jeans mass is high) and the cold neutral regions (where the
Jeans mass is low). During reionization, the IGM consists of these
two physically-independent phases which maintain their temperature
gap even as the filling factor of the ionized regions increases. This
temperature gap was not included in the semi-analytic models of
Shapiro, Giroux, \& Babul (1994) and Valageas \& Silk (1999), who
assumed a single-phase IGM with a uniform temperature. Gnedin \&
Ostriker (1997) also assumed a uniform photo-heating of the IGM in
their simulation.

A recent numerical simulation by Gnedin (1999) accounted for the
inhomogeneous distribution of the ionizing sources, with individual
ionization fronts surrounding each source. For a stellar spectrum,
most ionizing photons are just above the ionizing threshold of 13.6
eV, where the absorption cross-section is high and the mean-free-path
is correspondingly short. Therefore, the ionization fronts have a
sharp edge, and the reheating of the IGM occurs nearly simultaneously
with reionization (see Figure 3 in Gnedin 1999). Once reheating
occurs, the subsequent formation of low-mass galaxies is suppressed as
predicted by our semi-analytic model (see Figure 8 in Gnedin
1999). The suppression occurs more gradually in Gnedin (1999) than in
our model, because of a difference in the assumptions about star
formation. In the simulation of Gnedin (1999), gas which cools can
continue to form stars for much longer than a dynamical time, until
all of it has turned into stars. In our model, we assume that $10\%$
of the gas turns into stars, and that feedback heats or expels the
rest of the gas and thus halts star formation.  We assume that
additional star formation in the gas which is left over requires
another trigger such as a merger. Our assumption of an upper limit on
the efficiency of star formation appears to be required by
observations. Local galaxies contain a mass fraction in stars and
stellar remnants up to $\sim 1\%$ of the total halo mass (e.g.,
Fukugita, Hogan, \& Peebles 1998), which is an order of magnitude
smaller than the cosmic baryon fraction of $\ga 10\%$ indicated by
observations of X-ray clusters (White \& Fabian 1995; Arnaud \& Evrard
1999). Feedback is expected to have been most effective at limiting
star formation in the shallow potential wells which characterize high
redshift galaxies.

Chiu \& Ostriker (1999) included a two-phase IGM in their
semi-analytic model of reionization, and our results are generally
consistent with theirs. While we have focused on the star formation
rate and the faint luminosity function, Chiu \& Ostriker focused on
the reionization process itself and did not explicitly highlight the
suppression of star formation or the evolution of the luminosity
function.

\section{Conclusions}

We have shown that the reionization of the IGM is expected to have
left a clear signature on the history of the cosmic star formation
rate. In particular, the photo-ionization heating of the IGM
resulted in a suppression of the formation of low-mass galaxies, and
in a fairly rapid drop in the average SFR by a factor of $\sim 3$
relative to the case of no reionization.

Most of the additional star formation in the pre-reionization era occurred
in low-mass, faint galaxies. This means that {\it NGST} can detect only a
fraction of this additional star formation, but this should still suffice
to identify the reionization redshift. Even more striking is the
substantial increase (by 1--2 orders of magnitude) in the number density of
faint galaxies before reionization. These high-redshift galaxies are
expected to appear highly irregular because of their high merger
rates. Although the precise shape of the luminosity function depends on the
detailed model assumptions, the increase in the number density is dramatic
and robust. Much of this increase occurs at a flux limit $\la 1$ nJy and
should be detectable with an 8-meter {\it NGST}.

\acknowledgments

We thank Andrew Blain for providing SFR data from an earlier compilation.
We are also grateful to Jerry Ostriker, Nick Gnedin, and Weihsueh Chiu for
valuable discussions.  RB acknowledges support from Institute Funds. This work
was supported in part by NASA grants NAG 5-7039 and NAG 5-7768 for AL.




\begin{figure}
\plotone{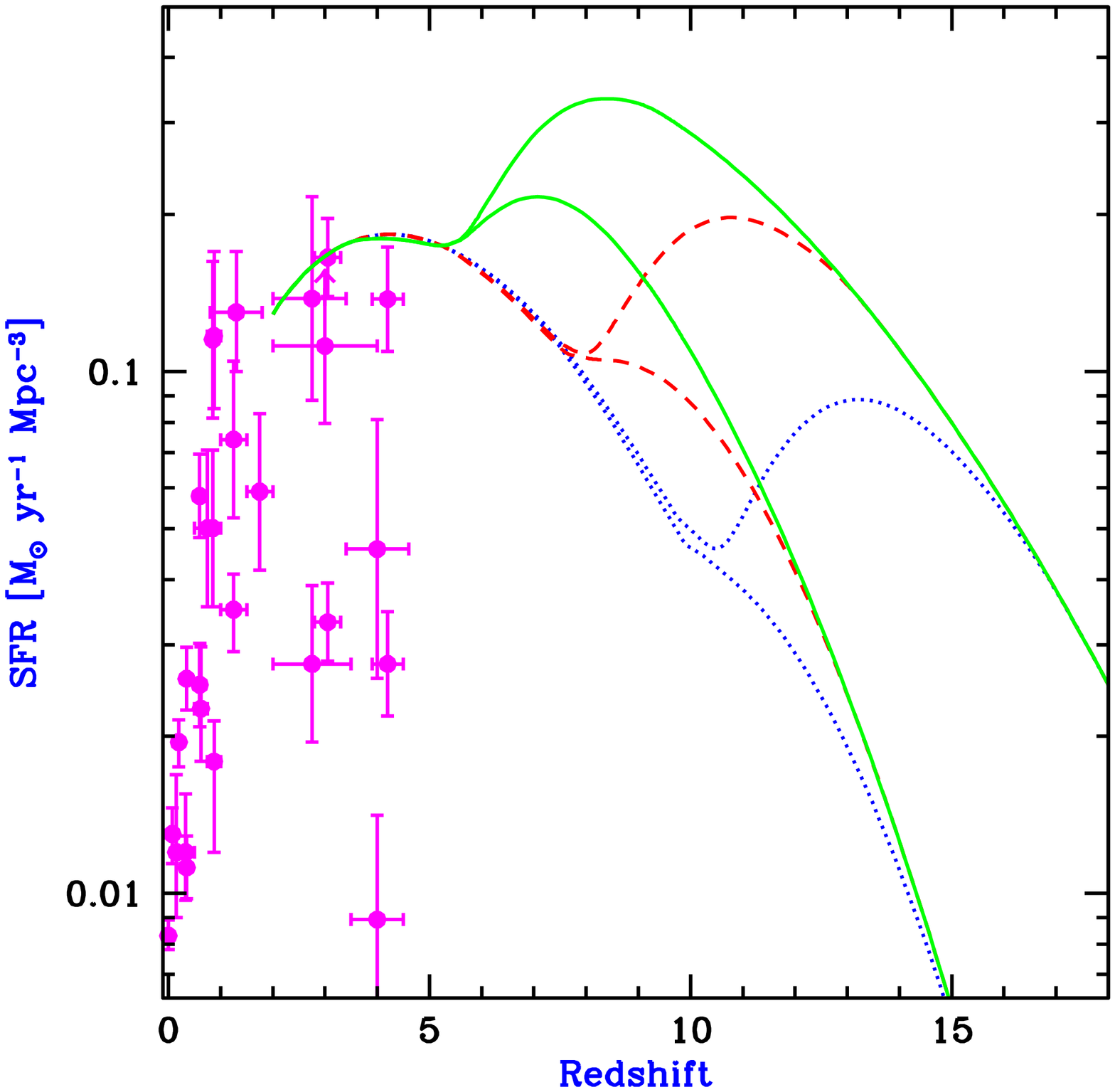}
\caption{Redshift evolution of the SFR (in $M_{\sun}$ per year per
comoving Mpc$^3$). Points with error bars are observational estimates
(compiled by Blain et al.\ 1999). Also shown are model predictions for
a reionization redshift $\zr=7$ (solid curves), $\zr=10$
(dashed curves), and $\zr=13$ (dotted curves), with a star
formation efficiency $\eta=10\%$. In each pair of curves, the upper
one is the total SFR, and the lower one is the fraction detectable
with {\it NGST} at a limiting point source flux of 0.25 nJy.}
\end{figure}

\begin{figure}
\plotone{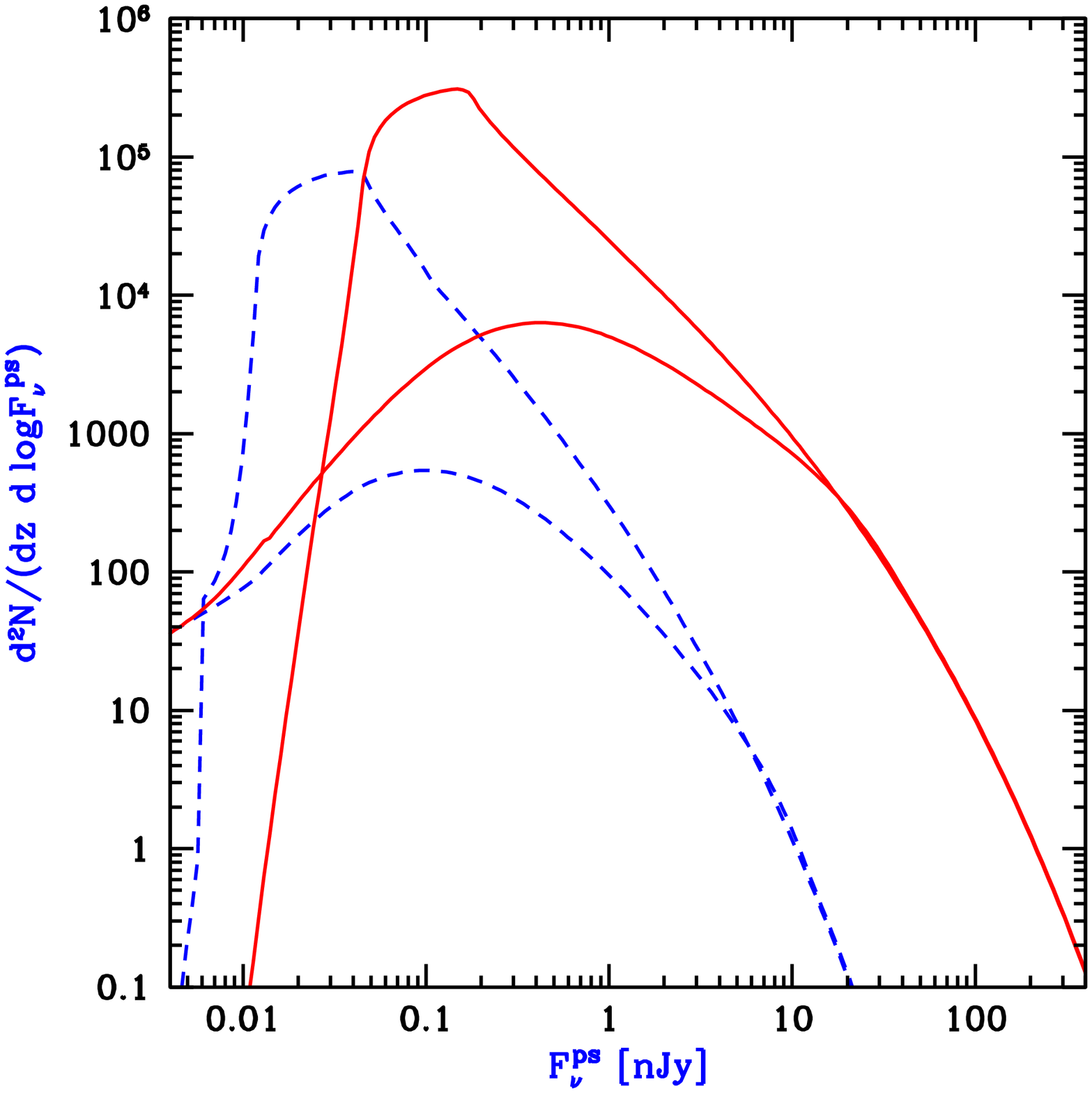}
\caption{Predicted luminosity function of galaxies at a fixed
redshift. With $\eta=10\%$, the curves show $d^2N/(dz\ d\log
F_{\nu}^{\rm ps})$, where $N$ is the total number of galaxies in a
single field of view of {\it NGST}, and $F_{\nu}^{\rm ps}$ is the
limiting point source flux at 0.6--3.5$\mu$m for {\it NGST}. Results
are shown at redshift $z=7$ (solid curves) or $z=13$ (dashed
curves). In each pair of curves, the upper one at 0.1 nJy assumes a
permanently neutral IGM (i.e., $\zr \ll z$), and the lower one assumes
a permanently fully-ionized IGM (i.e., $\zr \gg z$).}
\end{figure}


\begin{thebibliography}{bib0957}

\bibitem{clusters2}
Arnaud, M. \& Evrard, A. E. 1999, MNRAS, 305, 631
\bibitem{us2}
Barkana, R., \& Loeb, A. 1999a, ApJ, 523, 54
\bibitem{us}
Barkana, R., \& Loeb, A. 1999b, ApJ, in press (astro-ph/9901114)
\bibitem{Blain}
Blain, A. W., Jameson, A., Smail, I., Longair, M. S., Kneib, J.-P., \&
Ivison, R. J. 1999, MNRAS, submitted (astro-ph/9906311)
\bibitem{bryan}
Bryan, G., \& Norman, M. 1998, ApJ, 495, 80
\bibitem{8686} 
Chen, H-W., Lanzetta, K., \& Pascarelle, S. 1999, Nature, 398, 586 
\bibitem{Jerry}
Chiu, W., \& Ostriker, J. 1999, ApJ, submitted (astro-ph/9907220)
\bibitem{d98} 
Dey, A., Spinrad, H., Stern, D., Graham, J. R., \& Chaffee,
F. H. 1998, ApJ, 498, L93
\bibitem{Fan} 
Fan, X. et al. (SDSS collaboration) 1999, AJ, in press,
astro-ph/9903237
\bibitem{Fuk}
Fukugita, M., Hogan, C. J., \& Peebles, P. J. E. 1998, ApJ, 503, 518
\bibitem{Nick}
Gnedin, N. Y. 1999, ApJ, submitted (astro-ph/9909383)
\bibitem{go97} 
Gnedin, N. Y., \& Ostriker, J. P. 1997, ApJ, 486, 581
\bibitem{HAR}
Haiman, Z., Abel, T., \& Rees, M. 1999, ApJ, submitted (astro-ph/9903336)
\bibitem{HL97}
Haiman, Z., \& Loeb, A. 1997, ApJ, 483, 21
\bibitem{HL}
Haiman, Z., \& Loeb, A. 1998a, ApJ, 503, 505
\bibitem{HLb}
-----------------------------. 1998b, invited contribution to
Proc. of 9th Annual October Astrophysics Conference in Maryland, ``After
the Dark Ages: When Galaxies Were Young (the Universe at $2 < z < 5$)'',
College Park, October 1998 (astro-ph/9811395)
\bibitem{HL99}
Haiman, Z., \& Loeb, A. 1999, ApJ, 519, 479
\bibitem{HRL}
Haiman, Z., Rees, M., \& Loeb, A. 1996, ApJ, 476, 458;
erratum, ApJ, 484, 985
\bibitem{HTL}
Haiman, Z., Thoul, A. A., \& Loeb, A. 1996,
ApJ, 464, 523
\bibitem{hcm98} 
Hu, E. M., Cowie, L. L., \& McMahon, R. G. 1998, ApJ, 502, L99
\bibitem{KI}
Kitayama, T., \& Ikeuchi, S. 2000, ApJ, in press (astro-ph/9908084)
\bibitem{LC}
Lacey, C. G., \& Cole, S. M. 1993, MNRAS, 262, 627
\bibitem{Leith}
Leitherer et al. 1999, ApJS, 123, 3 
\bibitem{LR}
Loeb, A., \& Rybicki, G. B. 1999, ApJ, 524, 527
\bibitem{MHR}
Madau, P., Haardt, F., \& Rees, M. J. 1999, ApJ, 514, 648
\bibitem{MH}
Mihos, J.C. \& Hernquist, L. 1994, ApJ, 425, 13
\bibitem{MH2}
Mihos, J.C. \& Hernquist, L. 1996, ApJ, 464, 641
\bibitem{Jordi2}
Miralda-Escud\'e, J. 1998, ApJ, 501, 15
\bibitem{Jordi}
Miralda-Escud\'e, J., Haehnelt, M., \& Rees, M. 1999,  ApJ, submitted
(astro-ph/9812306)
\bibitem{Jordi3}
Miralda-Escud\'e, J., \&  Rees, M. 1998, ApJ, 497, 21
\bibitem{NFW}
Navarro, J. F., Frenk, C. S., \& White, S. D. M. 1997,
ApJ, 490, 493
\bibitem{NS}
Navarro, J. F., \& Steinmetz, M. 1997, ApJ, 478, 13
\bibitem{Peebles}
Peebles, P. J. E. 1993, Principles of Physical Cosmology
(Princeton: Princeton Univ.\ Press)
\bibitem{PS}
Press, W. H., \& Schechter, P. 1974, ApJ, 187, 425 
\bibitem{QKE}
Quinn, T., Katz, N., \& Efstathiou, G. 1996, MNRAS 278,
L49
\bibitem{Shapiro}
Shapiro, P. R., Giroux, M. L., \& Babul, A. 1994, ApJ, 427, 25
\bibitem{Shaver}
Shaver, P. A., Windhorst, R. A., Madau, P., \& de Bruyn, A. G. 1999,
\aap, 345, 380
\bibitem{Songaila}
Songaila, A., Hu, E. M., Cowie, L. L., \& McMahon, R. G. 1999,
ApJ, 525, L5
\bibitem{Spinrad}
Spinrad, H., Stern, D., Bunker, A., Dey, A., Lanzetta, K., Yahil,
A., Pascarelle, S., \& Fern\'{a}ndez-Soto, A. 1998, AJ, 116, 2617
\bibitem{sw67} 
Stecher, T. P., \& Williams, D. A. 1967, ApJ, 149, L29
\bibitem{TW}
Thoul, A. A., \& Weinberg, D. H. 1996, ApJ, 465, 608
\bibitem{silk}
Valageas, P., \& Silk, J. 1999, A\&A, in press (astro-ph/9903411)
\bibitem{NoWeymann}
Weymann, R. J., Stern, D., Bunker, A., Spinrad, H., Chaffee, F. H.,
Thompson, R. I., \& Storrie-Lombardi, L. J. 1998, ApJ, 505, 95
\bibitem{WHK}
Weinberg, D. H., Hernquist, L., \& Katz, N. 1997, ApJ, 477, 8
\bibitem{clusters}
White, D., \& Fabian, A. C. 1995, MNRAS, 273, 72
\bibitem{escape}
Wood, K., \& Loeb, A. 2000, ApJ, submitted (astro-ph/9911316)

\end{thebibliography}
\end{document}